\documentclass{SciPost}
\usepackage[bitstream-charter]{mathdesign}

\hypersetup{
    colorlinks,
    linkcolor={red!50!black},
    citecolor={blue!50!black},
    urlcolor={blue!80!black}
}
\DeclareSymbolFont{usualmathcal}{OMS}{cmsy}{m}{n}
\DeclareSymbolFontAlphabet{\mathcal}{usualmathcal}
\fancypagestyle{SPstyle}{
\fancyhf{}
\lhead{\colorbox{scipostblue}{\bf \color{white} ~SciPost Physics Core }}
\rhead{{\bf \color{scipostdeepblue} ~Submission }}

\fancyfoot[C]{\textbf{\thepage}}
}

\begin{document}

\pagestyle{SPstyle}

\begin{center}
{\Large \textbf{%
\color{scipostdeepblue}{
Boundary-induced switching of topological degeneracy in
the transverse field Ising chain.\\
}}}

\textbf{\ E. S. Ma \textsuperscript{} and Z. Song\textsuperscript{$\dagger$} 
}


\end{center}

\begin{center}
	{\bf } School of Physics, Nankai University, Tianjin 300071, China 
	\\[\baselineskip]
	$\dagger$ \href{mailto:email}{{\small songtc@nankai.edu.cn}} 

\end{center}

\section*{\color{scipostdeepblue}{Abstract}}

\textbf{\boldmath{The influence of boundary conditions on bulk properties is
generally expected to vanish in the thermodynamic limit. Here we show that
this expectation can fail in an open transverse field Ising chain with
modified boundary fields. By removing only the transverse fields acting on
the two edge spins while leaving the bulk Hamiltonian unchanged, we uncover
a boundary-induced switching of the topological spectral degeneracy. In
contrast to the conventional open chain, where the topological degeneracy
occurs in the ferromagnetic phase ($\left\vert g\right\vert <1$), the
modified boundary condition shifts it to the paramagnetic phase ($\left\vert
g\right\vert >1$). We demonstrate that the underlying mechanism originates
from two exact local conserved quantities associated with the edge spins,
which partition the Hilbert space into invariant sectors. Based on this
structure, we establish an exact description in terms of domain-wall
excitations, under which the Hamiltonian decomposes into two independent
Kitaev chains. Complementarily, the Majorana representation reveals that the
spectral switching originates from different unit-cell conventions and
boundary terminations of the corresponding effective SSH lattice, leading to
distinct boundary zero-mode structures without altering the bulk
Hamiltonian. Furthermore, we show that the boundary-induced topological
degeneracy remains observable at finite temperatures through non-Hermitian
quench dynamics, providing a dynamical manifestation of the bulk-boundary
correspondence. These results demonstrate that boundary engineering, through
the introduction of local conserved boundary degrees of freedom, provides an
effective route for controlling topological degeneracy in quantum spin
chains without modifying bulk interactions. }}

\vspace{\baselineskip}

\noindent%
\textcolor{white!90!black}{\fbox{\parbox{0.975\linewidth}{\textcolor{white!40!black}{\begin{tabular}{lr}  \begin{minipage}{0.6\textwidth}    {\small Copyright attribution to authors. \newline
    This work is a submission to SciPost Physics Core. \newline
    License information to appear upon publication. \newline
    Publication information to appear upon publication.}
  \end{minipage} & \begin{minipage}{0.4\textwidth}
    {\small Received Date \newline Accepted Date \newline Published Date}  \end{minipage}
\end{tabular}}
}}
} 


\vspace{10pt}
\noindent\rule{\textwidth}{1pt}
\tableofcontents
\noindent\rule{\textwidth}{1pt}
\vspace{10pt}

\section{Introduction}

\label{sec:intro} 
In conventional quantum many-body systems, the influence of boundary
conditions on macroscopic properties is generally expected to vanish in the
thermodynamic limit. By contrast, in topological systems, the bulk-boundary
correspondence establishes an intrinsic connection between bulk topology and
boundary excitations \cite%
{Hasan2010Colloquium,rudner2013Anomalous,graf2013bulk,Trifunovic2019higher}.
Over the past decade, extensive studies have demonstrated that boundary
termination \cite{Delplace2011,Malki2017tunable}, defects \cite%
{Teo2010topological,tang2025two}, and local edge manipulations can
profoundly modify boundary excitations and edge-state spectra without
altering the bulk Hamiltonian \cite%
{fu2008superconducting,Ezawa2018Minimal,Qi2011topological}, giving rise to a
variety of unconventional boundary phenomena, including Majorana zero modes 
\cite{kitaev2001unpaired,fu2008superconducting}, fractional boundary charges 
\cite{Benalcazar2019quantization,thakurathi2018from,park2016fractional}, and
higher-order corner or hinge states \cite%
{jin2023higher,peterson2018quantized,pu2023acoustic,choi2020evidence}. These
developments naturally raise the question of whether boundary manipulations
alone can also induce qualitative changes in the topological properties of
the entire many-body system.

Among exactly solvable quantum many-body models, the transverse field Ising
model (TFIM) occupies a central position in the study of quantum criticality 
\cite{fisher1995critical,curro2024quantum,Kinross2014evolution},
integrability \cite{Noh2021operator,peng2022bridging}, quantum many-body
scars \cite%
{van2020quantum,Yoshinaga2022emergence,peng2022bridging,corps2026scarred},
and quantum phase transitions \cite%
{sachdev2011quantum,dutta2015quantum,Hu2021first}. Through the Jordan-Wigner
transformation \cite{pfeuty1970one,sachdev2011quantum}, the one-dimensional
TFIM can be mapped onto a Kitaev chain supporting Majorana edge modes \cite%
{zhang2021quantum}, providing one of the simplest microscopic realizations
of the bulk-boundary correspondence. Owing to its exact solvability, the
TFIM has also become an important platform for investigating non-Hermitian
physics \cite{starkov2023quantum,du2023non,lu2024many}, dynamical quantum
phase transitions \cite{Heyl2013dynamical,Titum2019probing}, and
finite-temperature dynamics \cite%
{zhang2021quantum,zhang2022nonlocal,yang2022local}, underscoring its
enduring role as a paradigmatic model for exploring novel quantum many-body
phenomena. Experimentally, the transverse field Ising chain has been
realized in magnetic materials \cite{bitko1996quantum,coldea2010quantum}, as
well as in ultracold atoms trapped in optical lattices \cite%
{simon2011quantum,islam2011onset}.

Motivated by these developments, increasing attention has recently been
devoted to boundary effects in quantum spin systems. It has been shown that
boundary longitudinal fields in the Ising and Heisenberg chains can
substantially modify the excitation gap, edge modes, boundary dynamics, and
even induce boundary quantum phase transitions \cite%
{campostrini2015quantum,Hu2021first,Javed2025zeno,Pasnoori2023boundary,Javed2023counting,soto2026quantum}%
. These findings highlight boundary engineering as a powerful approach for
manipulating quantum many-body properties without altering the bulk
interactions. This naturally raises an intriguing question: Can a purely
local modification of the boundary transverse fields qualitatively change
the global physical properties of the system?

In this work, we address this question by investigating a modified open
boundary condition in which the transverse fields acting on the two edge
spins are removed; this boundary engineering gives rise to two exact local
conserved quantities associated with the edge spins. Remarkably, although
the bulk Hamiltonian remains unchanged, the parameter regime supporting
topological degeneracy is completely reversed: instead of appearing in the
ferromagnetic phase ($\left\vert g\right\vert <1$), the topological
degeneracy emerges in the paramagnetic phase ($\left\vert g\right\vert >1$).
To uncover the microscopic origin of this unexpected behavior, we develop
two complementary theoretical descriptions. First, we construct an exact
domain-wall representation, in which the Hamiltonian matrix in the
domain-wall basis decomposes into two identical copies of the Kitaev chain.
This correspondence naturally leads to a nonconventional criticality
indicator for the associated quantum phase transition. Second, we analyze
the model using the Jordan--Wigner transformation followed by the Majorana
representation, demonstrating that the reversal of the topological
degeneracy originates from different unit-cell conventions in the effective
SSH lattice, rather than from any change in the bulk topology.

Over the past few decades, the quantum phase transition in spin and
superconducting models has been extensively investigated through both
theoretical and experimental studies \cite%
{fisher1990quantum,bitko1996quantum,vojta2000quantum,si2001locally,
porras2004effective,uhlarz2004quantum,ronnow2005quantum,
coldea2010quantum,kim2011quantum,simon2011quantum,trenkwalder2016quantum,rem2019identifying}%
. Among the various approaches for probing quantum criticality, quantum
quench dynamics has emerged as a particularly powerful tool \cite%
{polkovnikov2011colloquium,essler2016quench,
abeling2016quantum,jurcevic2017direct, granet2020finite}. Following a sudden
quench, the Loschmidt echo (LE) \cite%
{andraschko2014dynamical,quan2006decay,cozzini2007quantum,Heyl2013dynamical,jafari2017loschmidt,mera2018dynamical}%
, which measures the overlap between the initial state and its time-evolved
counterpart, provides a sensitive probe of nonequilibrium dynamics \cite%
{polkovnikov2011colloquium}. By introducing a non-Hermitian perturbation and
studying the Loschmidt echo of thermal states, we show that the
boundary-induced topological degeneracy gives rise to qualitatively distinct
finite-temperature dynamics. These results establish a direct connection
between boundary engineering, bulk-boundary correspondence, and
nonequilibrium dynamics in quantum spin chains. Our work therefore provides
a simple yet conceptually transparent example illustrating how local
boundary manipulations can fundamentally reshape the topological properties
of an otherwise unchanged quantum many-body system.

The remainder of this paper is organized as follows. In Sec.~\ref{Model
Hamiltonians and spectral degeneracy}, we introduce the model Hamiltonians
with two types of open boundary conditions and present the concept of
spectral degeneracy. In Sec.~\ref{Domain-wall dynamics}, we develop a
domain-wall description of the model and establish its correspondence with
two independent Kitaev chains. In Sec.~\ref{PXP-type quantum criticality
indicator}, we propose a nonconventional quantum criticality indicator based
on the domain-wall representation. In Sec.~\ref{Majorana representation}, we
reexamine the model in the Majorana representation and elucidate the
underlying mechanism of the boundary-induced switching of the topological
degeneracy. In Sec.~\ref{Dynamical demonstration}, we investigate the
finite-temperature dynamics in the presence of non-Hermitian perturbations.
Finally, we summarize our results and discuss their implications in Sec.~\ref%
{conclusion}.

\section{Model Hamiltonians and spectral degeneracy}

\label{Model Hamiltonians and spectral degeneracy}

We begin with the 1D transverse field Ising model under different boundary
conditions. In general, most studies have focused on the model with periodic
boundary conditions. This is based on two considerations. First, the
translational symmetry enables the exact solution with the aid of the
Jordan-Wigner transformation and the Bogoliubov transformation in momentum
space. Second, people believe that boundary conditions cannot affect the
macroscopic properties in the thermodynamic limit. The corresponding
Hamiltonian takes the form%
\begin{equation}
H_{0}=-\sum_{j=1}^{N}\left( \sigma _{j}^{z}\sigma _{j+1}^{z}+g\sigma
_{j}^{x}\right) .  \label{H0}
\end{equation}%
Here $\sigma _{j}^{\alpha }$ $\left( \alpha =x,y,z\right) $ are the Pauli
operators on the $j$th site, satisfying the periodic boundary condition $%
\sigma _{N+1}^{\alpha }=\sigma _{1}^{\alpha }$. The parameter $g$ represents
the transverse field measured in units of the Ising interaction strength.
The exact solution reveals a quantum phase transition at $\left\vert
g\right\vert =1$, accompanied by a change in the ground-state degeneracy.
The ground state is twofold degenerate when $\left\vert g\right\vert <1$,
and becomes a singlet (non-degenerate) when $\left\vert g\right\vert >1$.

Under open boundary conditions, translational symmetry is broken.
Nevertheless, the model remains exactly solvable through the Jordan-Wigner
transformation followed by a real-space Bogoliubov transformation \cite%
{lieb1961two,dziarmaga2010dynamics}, and the energy level degeneracy can
still be analytically determined for the Hamiltonian%
\begin{equation}
H_{\text{\textrm{I}}}=H_{0}+\sigma _{1}^{z}\sigma _{N}^{z}.  \label{HI}
\end{equation}%
which removes the Ising interaction between the first and last spins from $%
H_{0}$. We refer to this as type I open boundary condition. Previous work
has shown that all the energy levels are twofold degenerate when $\left\vert
g\right\vert <1$. Importantly, this topological degeneracy, induced by open
boundary conditions, extends the quantum phase transition from the ground
state to excited states, persisting at finite temperatures.

We now turn to type-II boundary conditions. The corresponding Hamiltonian,%
\begin{equation}
H_{\text{\textrm{II}}}=H_{0}+\sigma _{1}^{z}\sigma _{N}^{z}+g\left( \sigma
_{1}^{x}+\sigma _{N}^{x}\right) ,  \label{HII}
\end{equation}%
is constructed by removing both the Ising coupling and transverse fields at
the boundary sites from $H_{0}$. Such a boundary condition can be regarded
as the idealized limit of site-dependent transverse fields. Recent advances
in programmable trapped-ion and Rydberg-atom quantum simulators enable
site-resolved control of local transverse fields through individual optical
addressing \cite%
{Levine2018high,scholl2021quantum,Cetina2022control,guo2024site}, making
this form of boundary engineering experimentally accessible in principle.
Intuitively, it should not affect the properties of the system so much.
However, as we shall demonstrate in the following sections, the situations
in the two regions are actually reversed. For clarity, the main conclusions
are summarized in Table~\ref{Table I} before proceeding to the detailed
analysis.

\begin{table}[tbp]
\caption{A summary of degeneracy properties under different boundary
conditions. Here, the related Hamiltonians have periodic boundary conditions
(PBC), Type I and Type II boundary conditions. The degeneracy occurs in the
ground state (GS), and in all eigenstates (All). The mechanisms of the
degeneracy for different cases are spontaneous symmetry breaking (SSB),
bulk-boundary correspondence (BBC), the $z$-component edge spin conservation
(ESC) [Eq.~(\protect\ref{ESC})], and global spin-flip symmetry (SFS) [Eq.~( 
\protect\ref{SFS})].}
\label{Table I}\centering
\par
\begin{tabular}{ccccc}
\hline\hline
Models & PBC & Type I & \multicolumn{2}{c}{Type II} \\ \hline
Regions & $\left\vert g\right\vert <1$ & $\left\vert g\right\vert <1$ & $%
\left\vert g\right\vert <1$ & $\left\vert g\right\vert >1$ \\ 
Degeneracy & GS & All & All & All \\ 
Fold & 2 & 2 & 2 & 4 \\ 
Mechanism & SSB & BBC & ESC+SFS & ESC+SFS, BBC \\ \hline\hline
\end{tabular}%
\end{table}

\section{Domain-wall dynamics}

\label{Domain-wall dynamics}

While an open Ising chain is typically investigated via the Majorana
representation of the Kitaev chain, this section focuses on the $H_{\text{ 
\textrm{II}}}$ model using an alternative method: a domain-wall excitation
picture. We will demonstrate that the spectrum of $H_{\text{\textrm{II}}}$
on an $N$-site chain decomposes into two identical copies of the Kitaev
chain on $(N-1)$-sites. Consequently, the results obtained for $H_{\text{ 
\textrm{I}}}$ carry over directly to $H_{\text{\textrm{II}}}$. The
underlying mechanism arises from the special boundary condition.

Under type-II boundary conditions, we have

\begin{equation}
\left[ \sigma _{1}^{z},H_{\text{\textrm{II}}}\right] =\left[ \sigma
_{N}^{z},H_{\text{\textrm{II}}}\right] =0,  \label{ESC}
\end{equation}%
this means that the boundary spins are frozen in eigenstates of $\sigma
_{1}^{z}$ and $\sigma _{N}^{z}$, respectively, as the transverse fields
vanish at the chain ends. In principle, there exist four invariant
subspaces, corresponding to the four possible configurations of the two end
spins. Here we restrict our analysis to two subspaces where the last spin is
fixed in the up and down states, respectively. Each of these subspaces
encompasses two of the four original subspaces, indexed by $\rho =\uparrow
,\downarrow $ according to the last spin state. The basis sets of the two
subspaces take the form 
\begin{equation}
\left\{ \left\vert \phi _{j}^{\uparrow }\right\rangle \right\}
:\prod_{l=1}^{N-1}\left\vert \sigma \right\rangle _{l}\left\vert \uparrow
\right\rangle _{N},  \label{subspace1}
\end{equation}%
and%
\begin{equation}
\left\{ \left\vert \phi _{j}^{\downarrow }\right\rangle \right\}
:\prod_{l=1}^{N-1}\left\vert \sigma \right\rangle _{l}\left\vert \downarrow
\right\rangle _{N},  \label{subspace2}
\end{equation}%
respectively, where $j=1,2,\dots,2^{N-1}$ and $\sigma=\uparrow ,\downarrow $%
. Here $\left\vert \uparrow \right\rangle _{l}$ and $\left\vert \downarrow
\right\rangle _{l}$ denote the eigenstates of $\sigma _{l}^{z}$ with
eigenvalues $+1$ and $-1$, respectively. Obviously, each basis state
consists of magnetic domains with different configurations, which can also
be used to represent the basis. We note that we have 
\begin{equation}
\left\{ \left\vert \phi _{j}^{\uparrow }\right\rangle \right\} =p\left\{
\left\vert \phi _{j}^{\downarrow }\right\rangle \right\} ,
\end{equation}%
where $p=\prod_{l=1}^{N}\sigma _{l}^{x}$ is the spin-flip operator or parity
operator. Consequently, all results obtained in one subspace are equally
applicable to the other, since $H_{\text{\textrm{II}}}$ possesses the
spin-flip symmetry%
\begin{equation}
\left[ p,H_{\text{\textrm{II}}}\right] =0.  \label{SFS}
\end{equation}

\begin{figure}[tbh]
	\centering
	\includegraphics[width=0.5\textwidth]{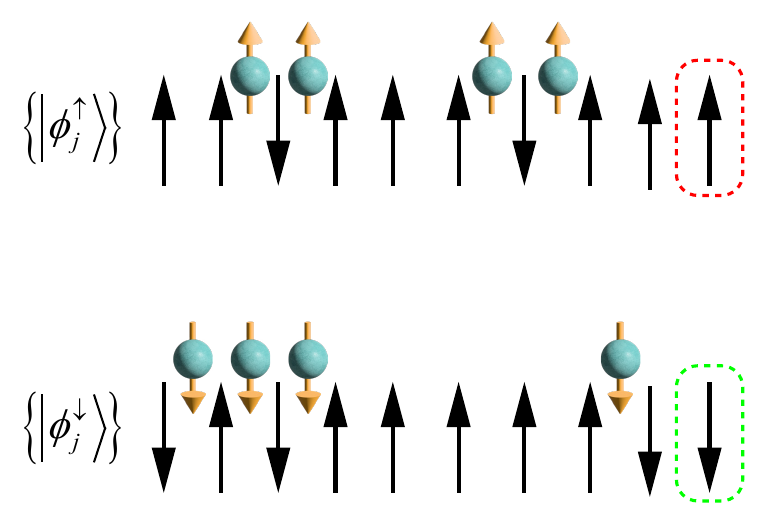}
	\caption{Schematic illustration of the mapping between spin states and
		fermionic states described by Eq.~(\protect\ref{mapping}). The spin states
		are divided into two subspaces where the last spin is in the spin-up state
		and spin-down state, respectively, as given by Eqs.~(\protect\ref{subspace1}
		) and (\protect\ref{subspace2}). Each subspace corresponds to a complete set
		of fermionic bases on $N-1$ sites, where each domain wall represents a
		fermionic excitation.}
	\label{fig1}
\end{figure}

In the following, we introduce a complete basis set based on the domain-wall
excitations. One can rewrite the basis through the following
transformations: 
\begin{equation}
\left\vert \uparrow \right\rangle _{l}\left\vert \downarrow \right\rangle
_{l+1}=\left\vert \downarrow \right\rangle _{l}\left\vert \uparrow
\right\rangle _{l+1}=\overline{\left\vert 1\right\rangle }_{l},\left\vert
\uparrow \right\rangle _{l}\left\vert \uparrow \right\rangle
_{l+1}=\left\vert \downarrow \right\rangle _{l}\left\vert \downarrow
\right\rangle _{l+1}=\overline{\left\vert 0\right\rangle }_{l},
\label{mapping}
\end{equation}%
where the states $\overline{\left\vert 1\right\rangle }_{l}$ and $\overline{%
\left\vert 0\right\rangle }_{l}$ are defined by 
\begin{equation}
c_{l,\rho }^{\dag }\overline{\left\vert 0\right\rangle }_{l}=\overline{%
\left\vert 1\right\rangle }_{l},\quad c_{l,\rho }\overline{\left\vert
0\right\rangle }_{l}=0.
\end{equation}%
Here $c_{l,\rho }^{\dag }$ ($c_{l,\rho }$) is the creation (annihilation)
operator of a spinful fermion. We note that there are two types of domain
walls and two types of magnetic domains. Both types of domain walls and both
types of magnetic domains are mapped to the same fermionic state and the
same vacuum state, respectively. While the duality of the mapping in Eq.~(%
\ref{mapping}) appears to suggest a bijective correspondence between the two
sets of bases, we can show that the mapping is in fact injective only within
the invariant subspace. Specifically, when the state of the last spin is
fixed, the type of the last domain wall is also fixed. The remaining domain
walls must then alternate between the two types, thereby determining the
types of the domains accordingly. Here, we give two examples for an $N=8$
lattice. They take the form%
\begin{equation}
\left\vert \uparrow \right\rangle _{1}\left\vert \downarrow \right\rangle
_{2}\left\vert \downarrow \right\rangle _{3}\left\vert \uparrow
\right\rangle _{4}\left\vert \uparrow \right\rangle _{5}\left\vert \uparrow
\right\rangle _{6}\left\vert \downarrow \right\rangle _{7}\left\vert
\uparrow \right\rangle _{8}=\overline{\left\vert 1\right\rangle }_{1}%
\overline{\left\vert 0\right\rangle }_{2}\overline{\left\vert 1\right\rangle 
}_{3}\overline{\left\vert 0\right\rangle }_{4}\overline{\left\vert
0\right\rangle }_{5}\overline{\left\vert 1\right\rangle }_{6}\overline{%
\left\vert 1\right\rangle }_{7},
\end{equation}%
and%
\begin{equation}
\left\vert \uparrow \right\rangle _{1}\left\vert \uparrow \right\rangle
_{2}\left\vert \downarrow \right\rangle _{3}\left\vert \downarrow
\right\rangle _{4}\left\vert \downarrow \right\rangle _{5}\left\vert
\uparrow \right\rangle _{6}\left\vert \downarrow \right\rangle
_{7}\left\vert \downarrow \right\rangle _{8}=\overline{\left\vert
0\right\rangle }_{1}\overline{\left\vert 1\right\rangle }_{2}\overline{%
\left\vert 0\right\rangle }_{3}\overline{\left\vert 0\right\rangle }_{4}%
\overline{\left\vert 1\right\rangle }_{5}\overline{\left\vert 1\right\rangle 
}_{6}\overline{\left\vert 0\right\rangle }_{7},
\end{equation}%
which belong to the two subspaces with $\rho =\uparrow ,\downarrow $,
respectively. Fig.~\ref{fig1} demonstrates the above correspondence.

In the following, we establish the correspondence between the spin and
fermionic operators by comparing their matrix elements in the two
representations. Without loss of generality, we perform the derivation in
the $\rho =\uparrow $ sector. As defined in Eq.~(\ref{subspace1}), the basis
in this sector is $\left\{ \left\vert \phi _{j}^{\uparrow }\right\rangle
\right\} $ and the corresponding domain-wall basis is 
\begin{equation}
\{|\psi _{j}^{\uparrow }\rangle \}:\prod_{l=1}^{N-2}\overline{|\tau \rangle }%
_{l},
\end{equation}%
where $\tau =0$ or $1$. Under the mapping introduced in Eq.~(\ref{mapping}),
the two basis states satisfy 
\begin{equation}
|\phi _{j}^{\uparrow }\rangle =|\psi _{j}^{\uparrow }\rangle .
\end{equation}%
Before proceeding, we introduce the projection operators 
\begin{equation}
P_{j}^{\sigma }=|\sigma \rangle _{j}\langle \sigma |_{j},
\end{equation}%
where $\sigma =\uparrow ,\downarrow $, together with the conventional ladder
operators 
\begin{equation}
\sigma _{j}^{\pm }=\frac{1}{2}\left( \sigma _{j}^{x}\pm i\sigma
_{j}^{y}\right) .  \label{ladder}
\end{equation}%
Furthermore, we introduce the following projected spin-flip operators: 
\begin{eqnarray}
P_{j,\mathrm{pair}} &=&P_{j-1}^{\uparrow }\sigma _{j}^{-}P_{j+1}^{\uparrow
}+P_{j-1}^{\downarrow }\sigma _{j}^{+}P_{j+1}^{\downarrow },  \nonumber \\
P_{j,\mathrm{hop}} &=&P_{j-1}^{\uparrow }\sigma _{j}^{-}P_{j+1}^{\downarrow
}+P_{j-1}^{\downarrow }\sigma _{j}^{+}P_{j+1}^{\uparrow }.
\end{eqnarray}%
According to Eq.~(\ref{mapping}), one readily verifies the following
relations: 
\begin{equation}
P_{k,\mathrm{pair}}\left\vert \phi _{l}^{\uparrow }\right\rangle
=c_{k-1}^{\dag }c_{k}^{\dag }\left\vert \psi _{l}^{\uparrow }\right\rangle
,P_{k,\mathrm{hop}}\left\vert \phi _{l}^{\uparrow }\right\rangle
=c_{k-1}^{\dag }c_{k}\left\vert \psi _{l}^{\uparrow }\right\rangle ,
\label{right}
\end{equation}%
which can be verified by enumerating all possible configurations of $\sigma
_{k-1}\sigma _{k}\sigma _{k+1}$ for $k\in \lbrack 2,N-1]$. Taking the
Hermitian conjugate of both sides yields 
\begin{equation}
\left\langle \phi _{j}^{\uparrow }\right\vert P_{k,\mathrm{pair}}^{\dag
}=\left\langle \psi _{j}^{\uparrow }\right\vert c_{k}c_{k-1},\left\langle
\phi _{j}^{\uparrow }\right\vert P_{k,\mathrm{hop}}^{\dag }=\left\langle
\psi _{j}^{\uparrow }\right\vert c_{k}^{\dag }c_{k-1},  \label{left}
\end{equation}%
Multiplying Eq.~(\ref{right}) from the left by $\langle \phi _{j}^{\uparrow
}|$ and Eq.~(\ref{left}) from the right by $|\phi _{l}^{\uparrow }\rangle $,
we obtain 
\begin{eqnarray}
\left\langle \phi _{j}^{\uparrow }\right\vert P_{k,\mathrm{pair}}\left\vert
\phi _{l}^{\uparrow }\right\rangle &=&\left\langle \psi _{j}^{\uparrow
}\right\vert c_{k-1}^{\dag }c_{k}^{\dag }\left\vert \psi _{l}^{\uparrow
}\right\rangle ,\left\langle \phi _{j}^{\uparrow }\right\vert P_{k,\mathrm{%
hop}}\left\vert \phi _{l}^{\uparrow }\right\rangle =\left\langle \psi
_{j}^{\uparrow }\right\vert c_{k-1}^{\dag }c_{k}\left\vert \psi
_{l}^{\uparrow }\right\rangle ,  \nonumber \\
\left\langle \phi _{j}^{\uparrow }\right\vert P_{k,\mathrm{pair}}^{\dag
}\left\vert \phi _{l}^{\uparrow }\right\rangle &=&\left\langle \psi
_{j}^{\uparrow }\right\vert c_{k}c_{k-1}\left\vert \psi _{l}^{\uparrow
}\right\rangle ,\left\langle \phi _{j}^{\uparrow }\right\vert P_{k,\mathrm{%
hop}}^{\dag }\left\vert \phi _{l}^{\uparrow }\right\rangle =\left\langle
\psi _{j}^{\uparrow }\right\vert c_{k}^{\dag }c_{k-1}\left\vert \psi
_{l}^{\uparrow }\right\rangle ,
\end{eqnarray}%
Summing the above four identities yields 
\begin{eqnarray}
&&\left\langle \phi _{j}^{\uparrow }\right\vert P_{k,\mathrm{pair}}+P_{k,%
\mathrm{hop}}+\mathrm{H.c.}\left\vert \phi _{l}^{\uparrow }\right\rangle
=\left\langle \phi _{j}^{\uparrow }\right\vert \sigma _{k}^{x}\left\vert
\phi _{l}^{\uparrow }\right\rangle  \nonumber \\
&=&\left\langle \psi _{j}^{\uparrow }\right\vert c_{k-1}^{\dag }c_{k}^{\dag
}+c_{k-1}^{\dag }c_{k}+\mathrm{H.c.}\left\vert \psi _{l}^{\uparrow
}\right\rangle =\left\langle \psi _{j}^{\uparrow }\right\vert \left(
c_{k-1}^{\dag }-c_{k-1}\right) \left( c_{k}^{\dag }+c_{k}\right) \left\vert
\psi _{l}^{\uparrow }\right\rangle .
\end{eqnarray}%
This demonstrates that $\sigma _{k}^{x}$ and $\left( c_{k-1}^{\dag
}-c_{k-1}\right) \left( c_{k}^{\dag }+c_{k}\right) $ reproduce identical
matrix elements in the two representations; the same conclusion also holds
in the $\rho =\downarrow $ sector. Therefore, we have the following
correspondence 
\begin{eqnarray}
P_{j-1}^{\uparrow }\sigma _{j}^{-}P_{j+1}^{\uparrow }+P_{j-1}^{\downarrow
}\sigma _{j}^{+}P_{j+1}^{\downarrow } &\mapsto &c_{j-1,\rho }^{\dag
}c_{j,\rho }^{\dag },  \nonumber \\
P_{j-1}^{\uparrow }\sigma _{j}^{-}P_{j+1}^{\downarrow }+P_{j-1}^{\downarrow
}\sigma _{j}^{+}P_{j+1}^{\uparrow } &\mapsto &c_{j-1,\rho }^{\dag }c_{j,\rho
},  \label{PXP}
\end{eqnarray}%
and 
\begin{equation}
\sigma _{j}^{x}\mapsto \left( c_{j-1,\rho }^{\dag }-c_{j-1,\rho }\right)
\left( c_{j,\rho }^{\dag }+c_{j,\rho }\right) .  \label{transverse term}
\end{equation}%
Applying the same procedure to the Ising interaction $\sigma _{j}^{z}\sigma
_{j+1}^{z}$, we obtain 
\begin{equation}
\sigma _{j}^{z}\sigma _{j+1}^{z}\mapsto 1-2c_{j,\rho }^{\dag }c_{j,\rho }.
\label{Ising term}
\end{equation}%
Based on these mappings, the original Ising Hamiltonian can be expressed as
two independent identical Kitaev chains, each of which takes the form%
\begin{equation}
H_{\text{\textrm{eq}}}^{[\rho ]}=-g\sum_{j=1}^{N-2}\left( c_{j,\rho }^{\dag
}c_{j+1,\rho }+c_{j,\rho }^{\dag }c_{j+1,\rho }^{\dag }\right) +\mathrm{H.c.}%
+\sum_{j=1}^{N-1}\left( 2c_{j,\rho }^{\dag }c_{j,\rho }-1\right) ,
\end{equation}%
with $\rho =\uparrow ,\downarrow $. Here, the mappings do not correspond to
the Jordan-Wigner transformation. To date, we have not found a
transformation relating the spin operators to the spinful fermionic
operators. The term \textquotedblleft equivalent\textquotedblright\ here
implies only that both Hamiltonians share the same spectrum---a conclusion
sufficient for our subsequent theoretical predictions. All the eigenstates
of the original Ising Hamiltonian have the topological degeneracy in the
region $\left\vert g\right\vert >1$. Across the entire parameter region, an
additional degeneracy emerges from the two independent identical Ising
chains. This degeneracy stems directly from the edge spin $z$-component
conservation and the spin-flip symmetry of $H_{\text{\textrm{II}}}$.
Notably, this situation is the reverse of that for the Ising chain under the
first type of boundary condition.

\begin{figure*}[tbh]
\centering
\includegraphics[width=1.0\textwidth]{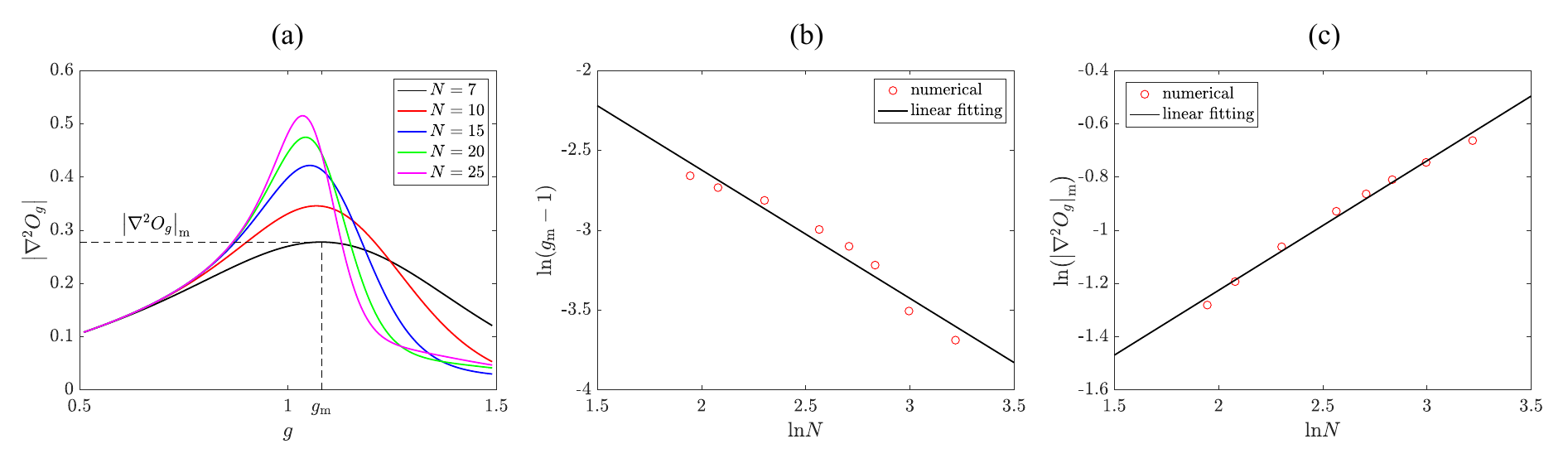}
\caption{(a) Absolute value of the second derivative of the order parameter
for different system sizes $N$. The order parameter and its corresponding
second derivative are defined in Eqs.~(\protect\ref{order}) and (\protect\ref%
{secondd}), respectively. The parameter $j_{0}$ is chosen at the center of
the chain. For $N=7, 10, 15, 20,$ and 25, the corresponding values of $j_{0} 
$ are 4, 5, 8, 10, and 13, respectively. $\left\vert \protect\nabla%
^{2}O_{g}\right\vert _{\mathrm{m}} $ denotes the maximum value of $%
\left\vert \protect\nabla ^{2}O_{g}\right\vert$, and $g_{\mathrm{m}}$ is the
corresponding value of $g$. (b) Plot of $\ln N$ versus $\ln\left( g_{ 
\mathrm{m}}-1\right) $. The red open circles represent the numerical results
for different system sizes $N$, and the black line is obtained from a linear
fit (c) Plot of $\ln N$ versus $\ln\left( \left\vert \protect\nabla %
^{2}O_{g}\right\vert _{\mathrm{m}}\right) $. The red open circles represent
the numerical results, and the black line is obtained from a linear fit.}
\label{fig2}
\end{figure*}

\section{PXP-type quantum criticality indicator}

\label{PXP-type quantum criticality indicator}

In the previous section, we analyzed $H_{\text{\textrm{II}}}$ from the
perspective of domain-wall excitations and established an exact
correspondence between the Hamiltonian matrix in the domain-wall basis and
that of two independent Kitaev chains within each invariant subspace. To the
best of our knowledge, such a domain-wall representation has not been
explicitly formulated for the present boundary-modified Ising chain. More
importantly, this construction is fundamentally different from the
conventional Jordan-Wigner transformation. The Jordan-Wigner transformation
is an exact operator mapping that transforms spin operators into fermionic
operators while preserving the operator algebra and commutation relations.
By contrast, the present approach is not an operator transformation.
Instead, it establishes a one-to-one correspondence between the basis states
of an invariant spin subspace and those of the corresponding domain-wall
Hilbert space, leading to an exact equivalence between the Hamiltonian
matrices rather than an equality between the spin and fermionic operators
themselves. Consequently, unlike the Jordan-Wigner transformation, the
correspondence established in Eqs.~(\ref{transverse term}) and (\ref{Ising
term}) is not an algebra-preserving operator transformation.

In this section, we demonstrate that the domain-wall representation not only
provides a microscopic understanding of the topological degeneracy, but also
offers an alternative bridge between the spin and fermionic descriptions,
enabling the identification of new physical observables.

In the one-dimensional Kitaev chain, the nearest-neighbor pairing
correlation \cite{Francica2022correlations,Medina2024correlation}, 
\begin{equation}
\left\langle \mathrm{G}\right\vert c_{j-1}^{\dag }c_{j}^{\dag }\left\vert 
\mathrm{G}\right\rangle ,
\end{equation}%
where $\left\vert \mathrm{G}\right\rangle $ is the ground state, provides a
useful diagnostic of the quantum phase transition, as its second derivative
with respect to the tuning parameter exhibits nonanalytic behavior at the
critical point. According to Eq.~(\ref{right}), the fermionic pair-creation
operator corresponds to a PXP-type operator in the spin representation---so
named after the PXP model \cite{Turner2018quantum}. Motivated by this
correspondence, we propose that 
\begin{equation}
O=\left\langle \mathrm{G}\right\vert P_{j_{0}-1}^{\uparrow }\sigma
_{j_{0}}^{-}P_{j_{0}+1}^{\uparrow }+P_{j_{0}-1}^{\downarrow }\sigma
_{j_{0}}^{+}P_{j_{0}+1}^{\downarrow }\left\vert \mathrm{G}\right\rangle ,
\label{order}
\end{equation}%
can serve as a quantum criticality indicator for detecting the quantum phase
transition in $H_{\text{\textrm{II}}}$. For finite systems, we calculate $O$
by exact diagonalization and evaluate the absolute value of its second
derivative with respect to the transverse field, 
\begin{equation}
\left\vert \nabla ^{2}O_{g}\right\vert =\left\vert \frac{\partial ^{2}O}{
\partial g^{2}}\right\vert ,  \label{secondd}
\end{equation}%
whose numerical results are presented in Fig.~\ref{fig2}(a). A pronounced
peak develops in the vicinity of the critical point $g=1$. As the system
size increases, the peak becomes progressively sharper and shifts closer to
the critical point, indicating the emergence of a singularity in the
thermodynamic limit.

To determine the asymptotic behavior as $N\rightarrow \infty $, we extract
the maximum value $\left\vert \nabla ^{2}O_{g}\right\vert _{ \mathrm{m}}$
together with its corresponding position $g_{\mathrm{m}}$, and perform a
finite-size scaling analysis, as shown in Figs.~\ref{fig2}(b) and \ref{fig2}%
(c). The fitted results indicate that $\left\vert \nabla
^{2}O_{g}\right\vert $ diverges at $g=1$ in the thermodynamic limit,
providing clear evidence of the quantum phase transition. Additional
numerical calculations further suggest that this indicator can also identify
quantum criticality in $H_{\text{\textrm{I}}}$, offering a complementary
perspective beyond conventional spin correlations \cite%
{Chitov2018local,pfeuty1970one,sachdev2011quantum}.

The transverse field Ising chain with next-nearest-neighbor interactions is
a natural extension of the standard model and has been extensively studied.
Interestingly, under the domain-wall representation, the
next-nearest-neighbor Ising interaction is mapped onto a nearest-neighbor
fermionic density-density interaction, 
\begin{equation}
\sigma _{j}^{z}\sigma _{j+2}^{z}\mapsto \left( 2c_{j,\rho }^{\dag }c_{j,\rho
}-1\right) \left( 2c_{j+1,\rho }^{\dag }c_{j+1,\rho }-1\right) ,
\end{equation}%
implying that $H_{\text{\textrm{II}}}$ with next-nearest-neighbor
interactions is equivalent to two identical interacting Kitaev chains with
nearest-neighbor density-density interactions. This correspondence provides
an alternative route for investigating interaction effects and many-body
phenomena in the Kitaev chain \cite{ma2026pairing}.

\section{\protect\bigskip Majorana representation}

\label{Majorana representation}

In this section, we reexamine the problem from an alternative perspective.
The same approach used for $H_{\text{\textrm{I}}}$ in Ref.~\cite%
{zhang2021quantum} can be applied to $H_{\text{\textrm{II}}}$. To obtain the
spectrum of $H_{\text{\textrm{II}}}$, one can map it onto a Kitaev model via
the Jordan-Wigner transformation as follows, 
\begin{equation}
c_{j}=\prod\limits_{i<j}\sigma _{i}^{x}\sigma _{j,(x)}^{+},\quad c_{j}^{\dag
}=\prod\limits_{i<j}\sigma _{i}^{x}\sigma _{j,(x)}^{-},
\end{equation}%
where $\sigma _{j,(x)}^{\pm }=\left( -\sigma _{j}^{z}\pm i\sigma
_{j}^{y}\right) /2$ corresponds to the ladder operators defined with respect
to the $x$-axis quantization convention, rather than the conventional $z$%
-axis quantization commonly adopted in the Jordan--Wigner transformation as
given in Eq.~(\ref{ladder}). This choice is made because the transverse
field in our model is applied along the $x$ direction. This transformation
maps a spin-down (spin-up) state to a fermionic state that is occupied
(empty). The inverse transformation is given by%
\begin{equation}
\sigma _{j}^{x}=1-2c_{j}^{\dag }c_{j},\sigma
_{j}^{z}=-\prod\limits_{i<j}\left( 1-2c_{i}^{\dag }c_{i}\right) \left(
c_{j}^{\dag }+c_{j}\right) .  \label{Jordan}
\end{equation}%
Substituting Eq.~(\ref{Jordan}) into $H_{\text{\textrm{II}}}$ yields 
\begin{equation}
H_{\text{\textrm{II}}}=-\sum_{j=1}^{N-1}\left( c_{j}^{\dag
}c_{j+1}+c_{j}^{\dag }c_{j+1}^{\dag }\right) +\mathrm{H.c.}%
+g\sum_{j=2}^{N-1}\left( 2c_{j}^{\dag }c_{j}-1\right) .  \label{Kitaev H}
\end{equation}%
Furthermore, it can be expressed in the Majorana representation 
\begin{equation}
H_{\text{\textrm{II}}}=-\frac{i}{2}\sum_{j=1}^{N-1}b_{j}a_{j+1}-\frac{i}{2}%
g\sum_{j=2}^{N-1}a_{j}b_{j}+0\times \left( a_{1}b_{1}+a_{N}b_{N}\right) +%
\text{\textrm{H.c.}},  \label{majorana}
\end{equation}%
by using the Majorana transformation 
\begin{equation}
a_{j}=c_{j}^{\dagger }+c_{j},\quad b_{j}=-i\left( c_{j}^{\dagger
}-c_{j}\right) ,
\end{equation}%
which satisfy the commutation relations 
\begin{equation}
\left\{ a_{j},a_{j^{\prime }}\right\} =\left\{ b_{j},b_{j^{\prime }}\right\}
=2\delta _{j,j^{\prime }},\quad \left\{ a_{j},b_{j^{\prime }}\right\} =0.
\end{equation}

\begin{figure}[tbh]
\centering
\includegraphics[width=0.6\textwidth]{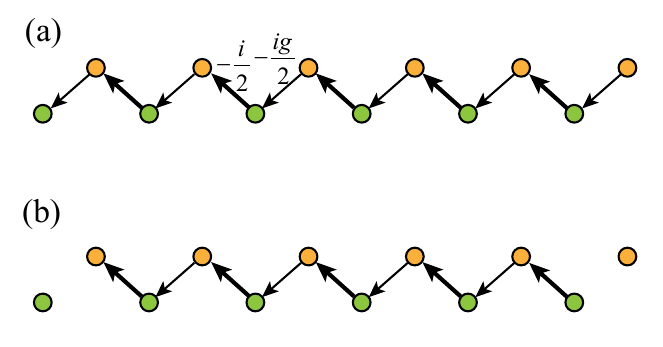}
\caption{Majorana lattice representations of $H_{\text{\textrm{I}}}$ (a) and 
$H_{\text{\textrm{II}}}$ of Eq.~(\protect\ref{Majorana}) (b). Unlike (a),
the chain in (b) exhibits two disconnected end sites, which induces exact
degeneracy in the spectrum and leads to an exchange of unit cells between
the two lattices. This exchange modifies the topological degeneracy.}
\label{fig3}
\end{figure}

To make the structure of the Majorana lattice clear, we write the
Hamiltonian in the basis $\varphi ^{T}=(a_{1},$ $b_{1},$ $a_{2},$ $b_{2},$ $%
a_{3},$ $b_{3},$ $\dots )$ as%
\begin{equation}
H_{\text{\textrm{II}}}=\varphi ^{T}h_{\text{\textrm{II}}}\varphi ,
\end{equation}%
where $h_{\text{\textrm{II}}}$ is a $2N\times 2N$ matrix. Explicitly,%
\begin{equation}
h_{\text{\textrm{II}}}=-\frac{i}{2}\left[ \sum\limits_{l=1}^{N-1}\left\vert
2l\right\rangle \left\langle 2l+1\right\vert
+\sum\limits_{l=2}^{N-1}g\left\vert 2l-1\right\rangle \left\langle
2l\right\vert \right] 0\times \left( \left\vert 1\right\rangle \left\langle
2\right\vert +\left\vert 2N-1\right\rangle \left\langle 2N\right\vert
\right) +\mathrm{H.c..}  \label{core matrix h}
\end{equation}%
Here $\left\{ \left\vert l\right\rangle ,\ l\in \lbrack 1,2N]\right\} $ is
an orthonormal basis satisfying $\left\langle l\right\vert l^{\prime
}\rangle =\delta _{ll^{\prime }}$. The structure of $h_{\text{\textrm{II}}}$
is schematically illustrated in Fig.~\ref{fig3}(b), which clearly forms a ($%
2N-2$)-site Su-Schrieffer-Heeger (SSH) chain with two isolated sites, $%
\left\vert 1\right\rangle \left\langle 1\right\vert $ and $\left\vert
2N\right\rangle \left\langle 2N\right\vert $, corresponding to the boundary
Majorana zero modes. For comparison, we also present the $2N$-site SSH chain
corresponding to the $H_{\text{\textrm{I}}}$ Hamiltonian. The physics of $H_{%
\text{\textrm{II}}}$ thus necessarily originates from that of a standard $%
(2N-2)$-site SSH chain, which is identical to that of $H_{\text{\textrm{I}}}$
with $g$ and $1/g$ exchanged. This agrees with the results presented in the
previous section.

Accordingly, based on the exact diagonalization of the SSH chain, $H_{\text{ 
\textrm{II}}}$ can be written in the diagonal form%
\begin{equation}
H_{\text{\textrm{II}}}=\sum_{n=1}^{N}\varepsilon _{n}\left( d_{n}^{\dagger
}d_{n}-\frac{1}{2}\right) ,
\end{equation}%
where the $d_{n}$ are fermionic operators satisfying standard
anticommutation relations. While the eigenvalues $\varepsilon _{n}$ can be
obtained numerically, closed-form analytical expressions are generally
unavailable for the SSH chain. Fortunately, there are two exact zero modes
in the large $N$ limit. The first one is%
\begin{equation}
d_{N}=\frac{c_{1}^{\dagger }+c_{1}+c_{N}^{\dagger }-c_{N}}{2},  \label{zero}
\end{equation}%
with $\varepsilon _{N}=0$, arising from the two isolated Majorana modes at
the chain ends. The second one is 
\begin{equation}
d_{N-1}=\Omega \sum_{j=1}^{N-1}g^{1-j}\left( c_{N-j+1}^{\dagger
}+c_{N-j+1}+c_{j}^{\dagger }-c_{j}\right) ,  \label{edge}
\end{equation}%
with $\varepsilon _{N-1}=0$ for $\left\vert g\right\vert >1$, where $\Omega =%
\sqrt{g^{2}-1}/2g$ is the normalization constant. Accordingly, we have 
\begin{equation}
\left[ d_{N}^{\dag },H_{\mathrm{II}}\right] =\left[ d_{N-1}^{\dag },H_{ 
\mathrm{II}}\right] =0,  \label{commutation}
\end{equation}%
which results in the 4-fold degeneracy for $\left\vert g\right\vert >1$, and
the 2-fold degeneracy for $\left\vert g\right\vert <1$. Indeed, for any
eigenstate $\left\vert \varphi \right\rangle $ of $H_{\mathrm{II}}$ with
energy $E$, the states $d_{N}\left\vert \varphi \right\rangle $, $%
d_{N-1}\left\vert \varphi \right\rangle $, $d_{N}^{\dag }\left\vert \varphi
\right\rangle $ and $d_{N-1}^{\dag }\left\vert \varphi \right\rangle $ are
also eigenstates with the same energy, provided they are nonzero.

These indicate that there are two types of degeneracy. The first one is the
degeneracy arising from edge spin $z$-component conservation given in Eq.~(%
\ref{ESC}) and the spin-flip symmetry given in Eq.~(\ref{SFS}), which is
equivalent to that from the operators $d_{N}$ and $d_{N}^{\dag }$. The
second one is a conditional degeneracy, related to the bulk-boundary
correspondence (BBC) of a Kitaev chain of length $N-1$. Specifically, the
existence of the operators $d_{N-1}$ and $d_{N-1}^{\dag }$ depends on the
existence of the zero modes in the corresponding SSH chain on a $2N-2$ site
lattice. Therefore, the degeneracy is topological since the zero modes are
robust. In comparison, the corresponding topological degeneracy for the type
I open boundary condition originates from the zero modes in the
corresponding SSH chain on a $2N$ site lattice. As shown in Fig.~\ref{fig3}, 
$2N$ and $2N-2$ SSH chains originate from the two different inter-unit-cell
cuttings from a ring system. In short, different open boundary conditions
lead to distinct effective SSH descriptions with distinct unit-cell
conventions and boundary terminations. Although the bulk topology remains
unchanged, different unit-cell conventions lead to different realizations of
the bulk-boundary correspondence and consequently distinct boundary
zero-mode structures. Specifically, for the conventional open boundary
condition shown in Fig.~\ref{fig3}(a), the edge zero modes appear in the
region $\left\vert g\right\vert <1$, whereas for the modified boundary
condition shown in Fig.~\ref{fig3}(b), they emerge in the region $\left\vert
g\right\vert >1$. 

\section{Dynamical demonstration}

\label{Dynamical demonstration}

The analysis in both the above sections shows that there are two types of
degeneracy. We focus on the observables related to the operators $d_{N-1}$
and $d_{N-1}^{\dag }$, which play a role in the dynamic demonstration of
such degeneracy.

Introducing a nonlocal operator 
\begin{equation}
D=-\frac{1}{2g}\sqrt{g^{2}-1}\sum_{j=1}^{N-1}g^{1-j}D_{j},
\end{equation}%
with a position-dependent component%
\begin{equation}
D_{j}=\prod\limits_{l<N+1-j}\left( \sigma _{l}^{x}\right) \sigma
_{N+1-j}^{z}+i\prod\limits_{l<j}\left( \sigma _{l}^{x}\right) \sigma
_{j}^{y},  \label{Dj}
\end{equation}%
we have the commutation relation 
\begin{equation}
\left[ D,H_{\mathrm{II}}\right] =\left[ D^{\dag },H_{\mathrm{II}}\right] =0,
\label{symmetry}
\end{equation}%
which is referred to as the edge-spin symmetry, since it is the outcome of
the edge operator of the Kitaev chain \cite{kitaev2001unpaired}. The
operator $D$ is a fermion operator obtained from the operator $d_{N-1}$,
obeying the relations $\left\{ D,D^{\dag }\right\} =1$ and $D^{2}=\left(
D^{\dag }\right) ^{2}=0$. It should be noted that it is contingent on the
following conditions: $\left\vert g\right\vert >1$, and the large $N$ limit.
In addition, the operator $D$ is non-universal and $g$-dependent.

Consequently, we have the following implications: the complete eigenstates $%
\left\{ \left\vert \psi _{n}^{+}\right\rangle ,\left\vert \psi
_{n}^{-}\right\rangle \right\} $ of $H_{\mathrm{II}}$ with eigen energy $%
E_{n}^{\pm }$, $H_{\mathrm{II}}\left\vert \psi _{n}^{\pm }\right\rangle
=E_{n}^{\pm }\left\vert \psi _{n}^{\pm }\right\rangle $, span two invariant
subspaces with 
\begin{equation}
p\left\vert \psi _{n}^{\pm }\right\rangle =\pm \left\vert \psi _{n}^{\pm
}\right\rangle ,
\end{equation}%
for any value of $g$. Importantly, within the region $\left\vert
g\right\vert >1$, we have the relations%
\begin{equation}
D\left\vert \psi _{n}^{+}\right\rangle =\left\vert \psi
_{n}^{-}\right\rangle ,\quad D^{\dag }\left\vert \psi _{n}^{-}\right\rangle
=\left\vert \psi _{n}^{+}\right\rangle ,D^{\dag }\left\vert \psi
_{n}^{+}\right\rangle =D\left\vert \psi _{n}^{-}\right\rangle =0,
\label{D_psi}
\end{equation}%
which guarantee the existence of degenerate eigenstates $%
E_{n}^{+}=E_{n}^{-}=E_{n}$, referred to as topological degeneracy
originating from the Majorana edge zero mode described by Eq.~(\ref{edge}).
Accordingly, we also have the relations 
\begin{equation}
\left( D^{\dag }-D\right) \left( \left\vert \psi _{n}^{+}\right\rangle \pm
\left\vert \psi _{n}^{-}\right\rangle \right) =\pm \left( \left\vert \psi
_{n}^{+}\right\rangle \mp \left\vert \psi _{n}^{-}\right\rangle \right) ,
\end{equation}%
which play an important role in the following.

Next, we employ quench dynamics to demonstrate our findings. Our strategy is
to examine the response of a thermal state to a nonlocal non-Hermitian
perturbation, which coalesces the topological degeneracy in the region $%
\left\vert g\right\vert >1$. It is expected that the dynamic responses for
initial thermal states in the two regions $\left\vert g\right\vert <1$ and $%
\left\vert g\right\vert >1$ are distinct.

Specifically, we consider a quench Hamiltonian by adding a non-Hermitian
perturbation, which is of the form 
\begin{equation}
\mathcal{H}=H_{\mathrm{II}}+\kappa D,
\end{equation}%
with $\kappa \ll g$. The motivation for taking such a perturbation is as
follows. In the region $\left\vert g\right\vert >1$, the relations in Eq.~(%
\ref{D_psi}) indicate that the matrix representation of $\mathcal{H}$ in
each subspace spanned by the two states $\left\vert \psi
_{n}^{+}\right\rangle $ and $\left\vert \psi _{n}^{-}\right\rangle $ is of
the form%
\begin{equation}
\mathcal{H}_{n}=\left( 
\begin{array}{cc}
E_{n} & 0 \\ 
\kappa & E_{n}%
\end{array}
\right) ,
\end{equation}%
which is a Jordan block. There is only one eigenstate $\left\vert \psi
_{n}^{ \mathrm{c}}\right\rangle =\left\vert \psi _{n}^{-}\right\rangle $,
which is referred to as the coalescing state. This makes the whole spectrum
of $H_{ \mathrm{II}}$ a coalescing spectrum, exhibiting fascinating dynamics
for any given initial state. Indeed, considering an arbitrary initial state $%
\left\vert \psi _{n}\left( 0\right) \right\rangle =c_{+}\left\vert \psi
_{n}^{+}\right\rangle +c_{-}\left\vert \psi _{n}^{-}\right\rangle $, the
evolved state is%
\begin{equation}
\left\vert \psi _{n}\left( t\right) \right\rangle =e^{-iE_{n}t}\left[
c_{+}\left\vert \psi _{n}^{+}\right\rangle +\left( c_{-}-it\kappa
c_{+}\right) \left\vert \psi _{n}^{-}\right\rangle \right] .
\end{equation}%
After a sufficiently long time, we have 
\begin{equation}
\left\vert \psi _{n}\left( \infty \right) \right\rangle \propto \left\vert
\psi _{n}^{\mathrm{c}}\right\rangle .  \label{EP}
\end{equation}%
This is an exclusive property of the exceptional point dynamics \cite%
{heiss2012physics, ma2025nonHermitian}. In contrast, in the region $g<1 $,
the dynamics is not sensitive to the perturbation. We then predict that the
final state always approaches the half component with fixed parity in the
region $\left\vert g\right\vert >1$, but remains almost unchanged in the
region $\left\vert g\right\vert <1$.

Below, we take the thermal equilibrium state of $H_{\mathrm{II}}$ as the
initial state, which can be described by the density matrix 
\begin{equation}
\rho \left( 0\right) =\frac{e^{-\beta H_{\mathrm{II}}}}{\mathrm{Tr}e^{-\beta
H_{\mathrm{II}}}},
\end{equation}%
where $\beta =1/T$ is the reciprocal of the temperature. Under the driving
of $\mathcal{H}$, the density matrix obeys the equation 
\begin{equation}
i\frac{\partial \rho \left( t\right) }{\partial t}=\mathcal{H}\rho \left(
t\right) -\rho \left( t\right) \mathcal{H}^{\dag }.
\end{equation}%
Therefore, we have the formal solution 
\begin{equation}
\rho \left( t\right) =e^{-i\mathcal{H}t}\rho \left( 0\right) e^{i\mathcal{H}
^{\dag }t},
\end{equation}%
which preserves Hermiticity and positive semidefiniteness throughout the
evolution. To ensure that the trace of $\rho \left( t\right) $ remains $1$,
following the standard treatment of non-Hermitian quantum dynamics \cite%
{brody2014biorthogonal,Gopalakrishnan2021}, we normalize $\rho \left(
t\right) $ by setting 
\begin{equation}
\rho \left( t\right) =e^{-i\mathcal{H}t}\rho \left( 0\right) e^{i\mathcal{H}
^{\dag }t}/\mathrm{Tr}\left[ e^{-i\mathcal{H}t}\rho \left( 0\right) e^{i 
\mathcal{H}^{\dag }t}\right] .
\end{equation}
Consequently, the normalized density matrix remains a proper density matrix
with unit trace. We then introduce the generalized Loschmidt echo (LE),
which is essentially the Uhlmann fidelity \cite{uhlmann1976transition}, to
characterize the fidelity between the evolved state and the initial state,
which is defined by 
\begin{equation}
L\left( t\right) =\left[ \mathrm{Tr}\sqrt{\sqrt{\rho \left( 0\right) }\rho
\left( t\right) \sqrt{\rho \left( 0\right) }}\right] ^{2}.  \label{LE}
\end{equation}%
In the region $\left\vert g\right\vert >1$, after a long time evolution,
only the coalescing component in the subspace $\left\{ \left\vert \psi
_{n}^{ \mathrm{c}}\right\rangle \right\} $ survives, and then the Loschmidt
echo approaches $0.5$. However, in the region $\left\vert g\right\vert <1$,
the non-Hermitian perturbation does not dramatically affect the energy
levels, and the Loschmidt echo is expected to remain close to $1$.

\begin{figure}[tbh]
	\centering
	\includegraphics[width=0.6\textwidth]{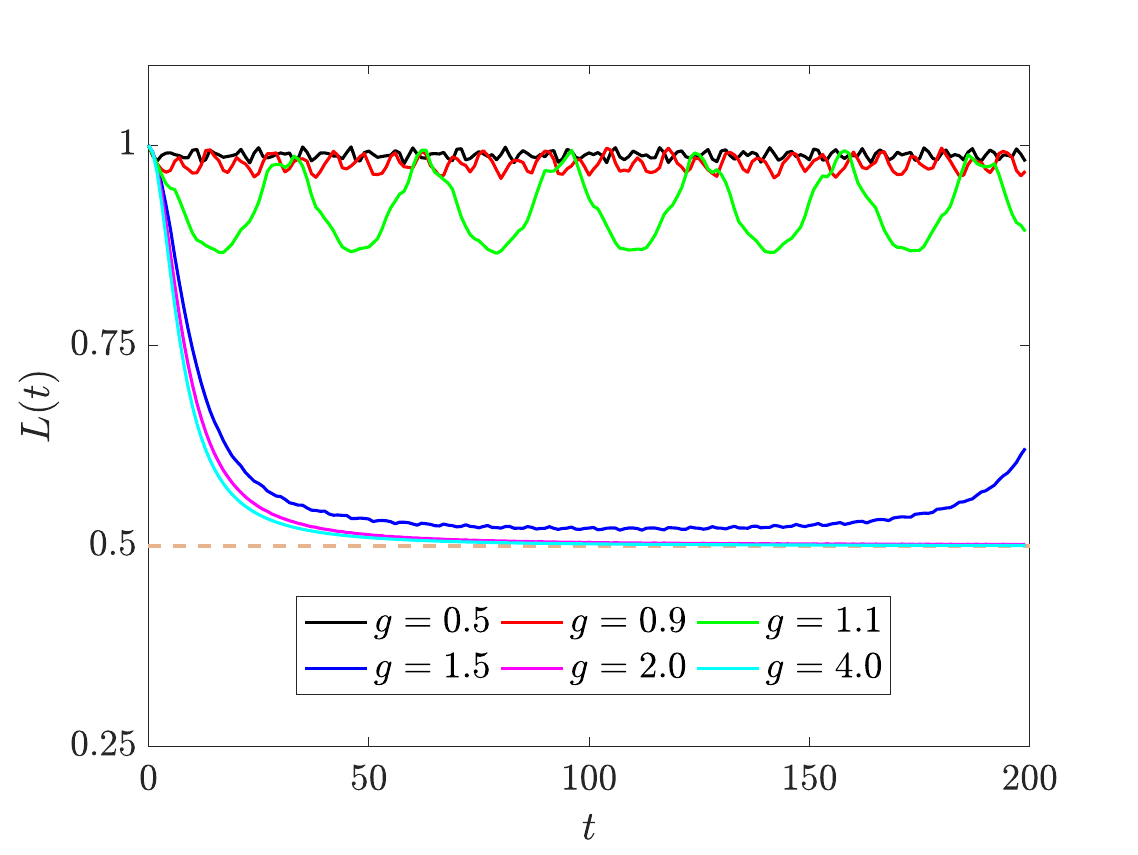}
	\caption{Plots of the Loschmidt echo as defined in Eq.~(\protect\ref{LE})
		for several representative transverse field values. The driving Hamiltonian
		is $\mathcal{H}=H_{\mathrm{II}}+\protect\kappa D_{1}$ and numerical
		simulations are performed independently in subspaces spanned by $\left\{
		\left\vert \protect\phi _{j}^{\uparrow }\right\rangle \right\}$ and $\left\{
		\left\vert \protect\phi _{j}^{\downarrow }\right\rangle \right\}$. Each
		subspace gives identical results. Other parameters are $\protect\beta =2$, $%
		\protect\kappa=0.1$, and $N=11$.}
	\label{fig4}
\end{figure}

The exact EP dynamics is based on the non-Hermitian perturbation $D$;
however, $D$ is a linear combination of $D_{j}$ with power-law decaying
weights, which means that $D_{1}$ is the leading term. As a result, we take $%
\mathcal{H}=H_{\mathrm{II}}+\kappa D_{1}$ as the quench Hamiltonian; then
the dynamics does not substantially deviate from the exact EP dynamics. It
is worth noting that $D_{j}$ preserves the two invariant subspaces $\left\{
\left\vert \phi _{j}^{\uparrow }\right\rangle \right\}$ and $\left\{
\left\vert \phi _{j}^{\downarrow }\right\rangle \right\}$ given by Eqs.~(\ref%
{subspace1}) and (\ref{subspace2}). Consequently, numerical simulations can
be performed in each subspace independently, with both expected to yield
identical results. The numerical results for several representative $g$
under finite temperature are presented in Fig.~\ref{fig4}, which accord with
the analytical analyses.

This detection scheme exploits a distinctive feature of non-Hermitian
exceptional point dynamics described by Eq.~(\ref{EP}); its effectiveness
relies on the commutation relations in Eq.~(\ref{symmetry}), which originate
from the boundary zero modes. In the thermodynamic limit, these relations
hold throughout the entire topological region $\left\vert g\right\vert >1$.
For finite systems, however, finite-size effects slightly lift the
topological degeneracy, so a larger value of $\left\vert g\right\vert$ is
required for the boundary zero modes to become well localized. As shown in
Fig.~\ref{fig4}, the Loschmidt echo approaches the theoretical value of $0.5$
only for $g\geq 2$. Therefore, although the operator $D$ is $g$-dependent,
the physical conclusion---that the boundary-induced topological degeneracy
persists throughout the entire $\left\vert g\right\vert >1$ phase in the
thermodynamic limit---is universal. By contrast, for a generic Hermitian
perturbation, the two degenerate states $\left\vert \psi _{n}^{\pm
}\right\rangle$ remain orthogonal and coexist throughout the time evolution.
Consequently, no exceptional-point dynamics occurs, so the Loschmidt echo
remains unity for all values of $g$; therefore, a generic Hermitian
perturbation cannot serve as an effective dynamical probe of the topological
degeneracy. Thus, the purpose of introducing the non-Hermitian perturbation
is not to establish a universal dynamical response, but rather to provide a
convenient theoretical method for identifying the boundary-induced
topological degeneracy.

\begin{figure}[tbh]
	\centering
	\includegraphics[width=1.0\textwidth]{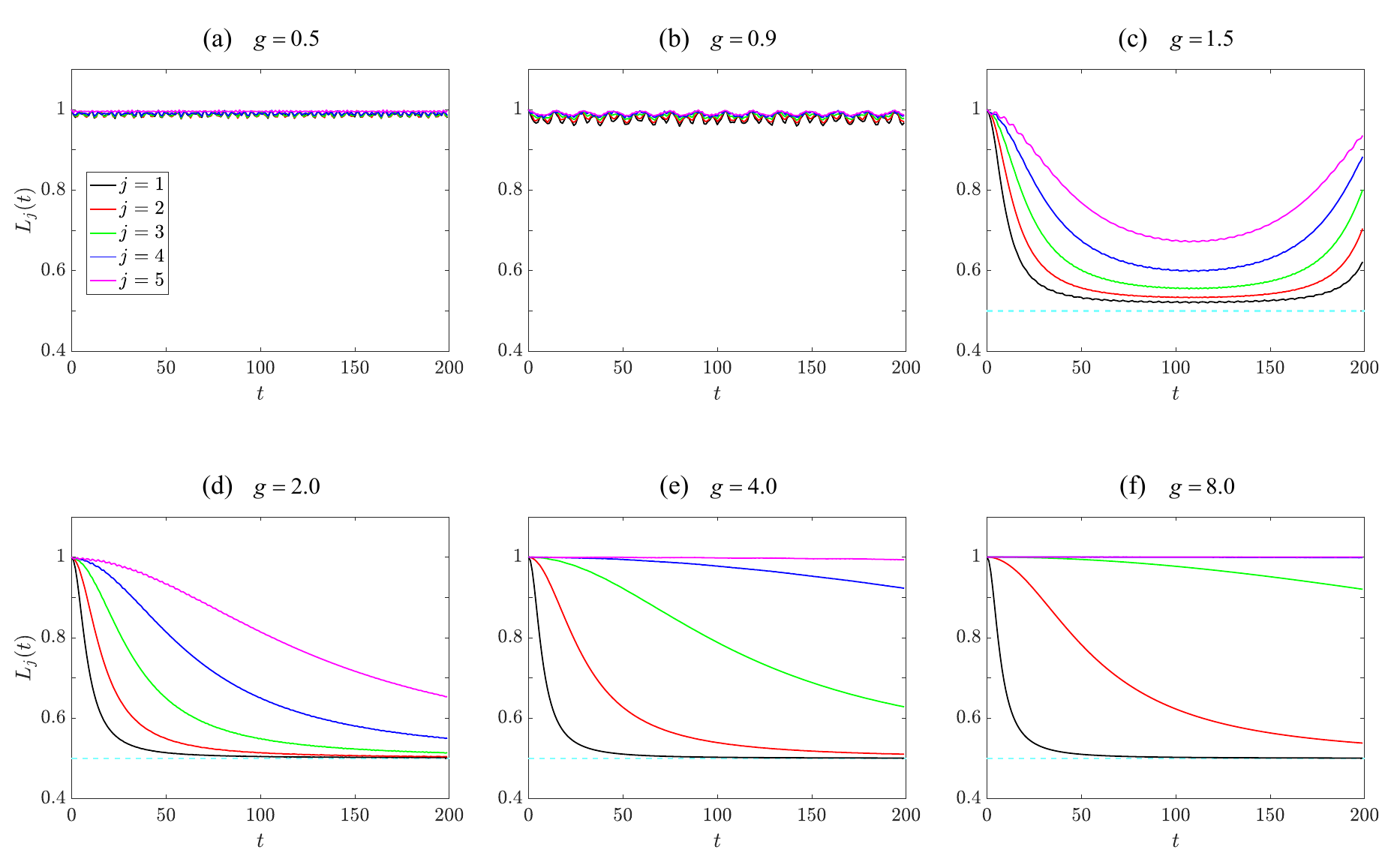}
	\caption{Position-dependent Loschmidt echo for several representative values
		of the parameter $g$ following a quench described by Eq.~(\protect\ref{Lj}).
		Time evolution is performed independently in the two invariant subspaces
		[Eqs.~(\protect\ref{subspace1}) and (\protect\ref{subspace2})], yielding
		identical results. Other parameters are $\protect\beta =2$, $\protect\kappa %
		=0.1 $, and $N=11$.}
	\label{fig5}
\end{figure}

In the following, we delve into the bulk-boundary correspondence at finite
temperature through quench dynamics. Considering the following quench
Hamiltonian 
\begin{equation}
\mathcal{H}=H+\kappa D_{j},  \label{Lj}
\end{equation}%
and the corresponding Loschmidt echo is denoted $L_{j}\left( t\right) $. In
Fig.~\ref{fig5} we show the numerical simulations for different phases;
these results indicate that for $g<1$ all $L_{j}\left( t\right) $ remain
close to $1$; however, for $g>1$ the marginal Loschmidt echoes converge to $%
0.5$ after a sufficiently long time and the closer $j$ is to the end the
more rapidly $L_{j}\left( t\right) $ approaches $0.5$. In addition, due to
the finite system size, the larger $g$ is, the more obvious the marginal
effects become.

Before concluding this work, we would like to emphasize that our work does
not concern the conventional boundary-induced degeneracy of topological
phases. Instead, we reveal a boundary-condition-driven quantum phase
switching in an interacting Ising system, which is characterized by an
unconventional order parameter. Importantly, this phase switching is
manifested not merely through a change in the degeneracy, but through a
reconstruction of the many-body eigenstates across the phase boundary. To
make this distinction clear, we analyze the evolution of the energy levels
as a function of the boundary condition, which explicitly demonstrates the
reconstruction of the eigenstates and the associated quantum phase switching.

\begin{figure*}[tbh]
	\centering
	\includegraphics[width=0.9\textwidth]{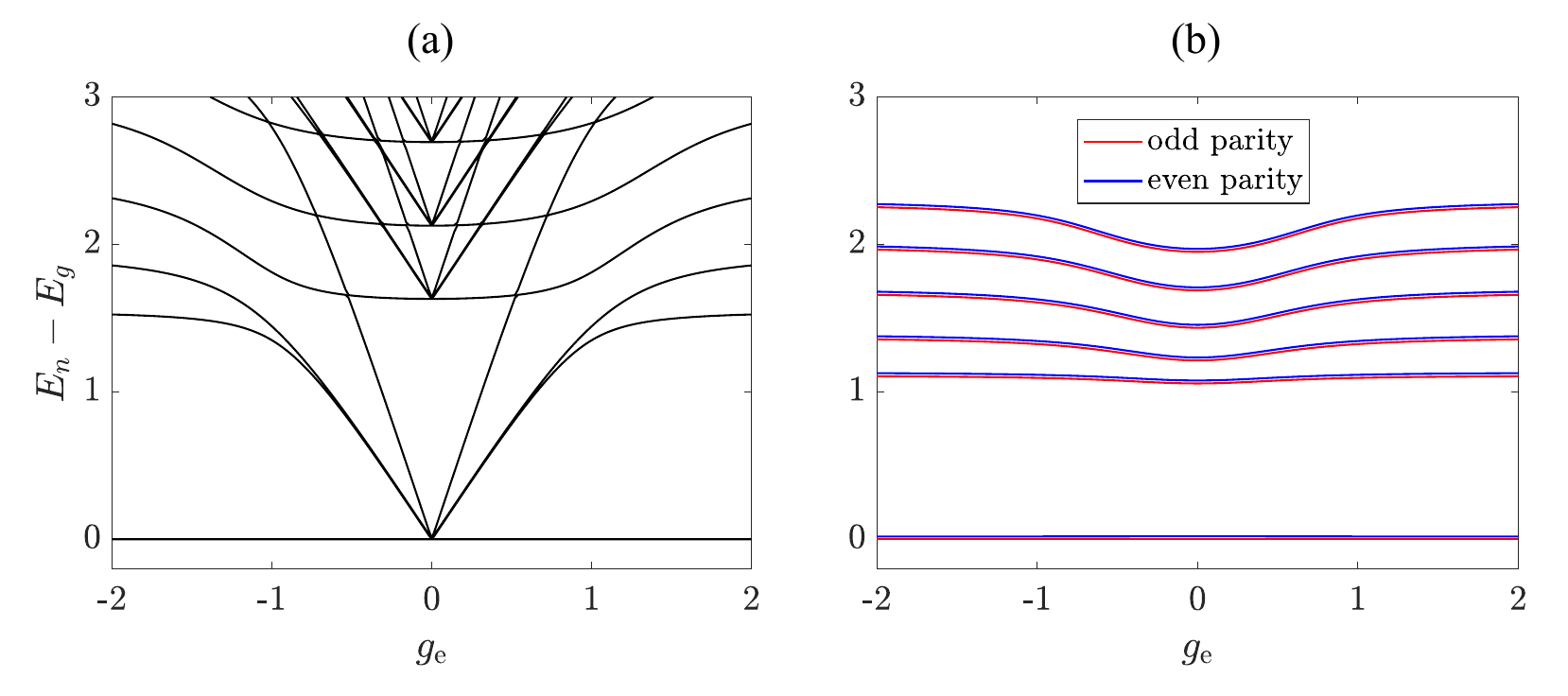}
	\caption{ Several low-energy levels of the open Ising chain described by
		Eq.~(\protect\ref{H_e}). Here, $E_{g}$ and $E_{n}$ denote the ground-state
		energy and the $n$th excited-state energy, respectively. (a) Spectrum of $%
		H(g,g_{\mathrm{e}})$ for $g=1.7$ and $N=12$. (b) Spectrum of $H(g,g_{\mathrm{%
				e}})+0.01p$, where $p=\prod_{l=1}^{N}\protect\sigma _{l}^{x}$ is the parity
		operator. The other parameters are $g=0.5$ and $N=12$. The perturbation $%
		0.01p$ lifts the twofold degeneracy between states with odd and even parity,
		represented by the red and blue lines, respectively.}
	\label{fig6}
\end{figure*}

We consider the following Hamiltonian, 
\begin{equation}
H(g,g_{\mathrm{e}})=-\sum_{j=1}^{N-1}\sigma _{j}^{z}\sigma
_{j+1}^{z}-g\sum_{j=2}^{N-1}\sigma _{j}^{x}-g_{\mathrm{e}}(\sigma
_{1}^{x}+\sigma _{N}^{x}),  \label{H_e}
\end{equation}%
where $g$ and $g_{\mathrm{e}}$ denote the bulk and edge transverse fields,
respectively. This Hamiltonian interpolates between the type-I and type-II
boundary conditions by tuning $g_{\mathrm{e}}$: specifically, $H(g,g_{ 
\mathrm{e}})=H_{\mathrm{I}}$ for $g_{\mathrm{e}}=g$ and $H(g,g_{\mathrm{e}%
})=H_{\mathrm{II}}$ for $g_{\mathrm{e}}=0$. In Fig.~\ref{fig6}, for a finite
system with fixed $g$, we present the energy-level differences between the
ground state and the excited states as $g_{\mathrm{e}}$ varies over the
interval $[-2,2]$. These results show that the ground state is always
two-fold degenerate for finite $g_{\mathrm{e}}$ in the regime $|g|<1,$ and
that this degeneracy can be lifted by introducing a specific perturbation.
In contrast, for $|g|>1$, the ground state is non-degenerate for finite $g_{%
\mathrm{e}}$, except at $g_{\mathrm{e}}=0$, where it becomes four-fold
degenerate. In the thermodynamic limit, it can be shown that any finite $g_{%
\mathrm{e}}$ yields the same topological phase diagram as the type-I
boundary condition in Eq.~{(\ref{HI})}. In contrast, $g_{\mathrm{e}}=0 $
constitutes a nontrivial boundary point that switches the regime supporting
zero modes. This reveals that specially engineered boundary conditions can
induce a quantum phase switching.

\section{Conclusion}

\label{conclusion}

In this work, we have investigated the role of boundary engineering in the
transverse field Ising chain by comparing two different types of open
boundary conditions. Although the two models differ only by the transverse
fields acting on the boundary spins, they exhibit qualitatively different
topological degeneracies. Remarkably, removing the boundary transverse
fields reverses the parameter region supporting the topological degeneracy
without modifying any bulk interaction.

To elucidate the microscopic origin of this phenomenon, we developed two
complementary theoretical descriptions. We first established a domain-wall
representation based on two exact local conserved quantities associated with
the edge spins, under which the Hamiltonian is shown to be exactly
equivalent, at the Hamiltonian-matrix level, to two independent Kitaev
chains. Besides providing an intuitive physical picture beyond the
conventional Jordan-Wigner transformation, this representation offers an
alternative correspondence between spin and fermionic operators.
Complementarily, the Jordan-Wigner transformation followed by the Majorana
representation reveals that the observed switching originates from different
unit-cell conventions and boundary terminations of the effective SSH
lattice, giving rise to distinct boundary zero-mode structures while
preserving the bulk Hamiltonian. We further demonstrated that the
boundary-induced topological degeneracy remains observable at finite
temperature through non-Hermitian quench dynamics, providing a dynamical
manifestation of the bulk-boundary correspondence.

Our results demonstrate that introducing local conserved boundary degrees of
freedom through boundary engineering provides an effective route for
controlling topological properties in quantum spin chains without modifying
bulk interactions.

\section*{Acknowledgements}


\paragraph{Funding information}

This work was supported by the National Natural Science Foundation of China
(under Grant No. 12374461).






\bibliography{Ising-Kitaev.bib}


\end{document}